\documentstyle[12pt,epsfig]{article}
\begin{document}
\begin{center}
{\bf Large-Order Perturbation Theory in Infrared-Unstable\\
Superrenormalizable Field Theories}\\[.15in]
John M. Cornwall*\\[.15in]
{\it Physics Department, University of California\\
405 S. Hilgard Ave., Los Angeles Ca 90095-1547}\\[.15in]
{\bf Abstract}
\end{center}
We study the factorial divergences of Euclidean $\phi^3_5$, a problem with connections both to high-energy multiparticle scattering in d=4 and to d=3 (or high-temperature) gauge theory, which like $\phi^3_5$ is infrared-unstable and superrenormalizable. At large external momentum $p$ (or small mass $M$) and large order $N$ one might expect perturbative bare skeleton graphs to behave roughly like $N!(ag^2/p)^N$ with $a>0$, so that no matter how large $p$ is there is an $N\sim p/g^2$ giving rise to strong perturbative amplitudes.  The semiclassical Lipatov technique (which only works in the presence of a mass) is blind to this
momentum dependence, so we proceed by direct summation of bare skeleton graphs.  We find that the various limits of large $N$, large $p$, and small mass $M$ do not commute, and that when $N\gg p^2/M^2$ there is a Borel singularity associated with $g^2/M$, not $g^2/p$.  This is described by the zero-momentum Lipatov technique, and we find the necessary soliton for $\phi^3_5$; the corresponding sphaleron-like solution for unbroken Yang-Mills (YM) theory has long been known. We also show that the massless theories have no classical solitons.  We discuss non-perturbative effects based partly on known physical arguments concerning the cancellation by solitons of imaginary parts due to the perturbative Borel singularity, and partly on the dressing of bare skeleton graphs by dressed propagators showing non-perturbative mass generation, as happens in d=3 gauge theory. \\[12pt]
UCLA/96/TEP/36 \mbox{} \hfill December 1996\\
\footnoterule
\noindent *Electronic addresss:  Cornwall@physics.ucla.edu
\newpage
\section{Introduction}      
Several interesting questions arise about the large-order behavior of
perturbation theory in (Euclidean) infrared-unstable theories like d=3 QCD or electroweak (EW) $SU(2)$, which have
dimensionful coupling constants and are superrenormalizable.  These theories have physical interest in themselves as the $N=0$ Matsubara-mode magnetic sector of high-temperature (above the phase transition) gauge theories, which is strongly-coupled even for EW theory, and there is a connection also to the behavior of large-multiplicity high-energy scattering amplitudes.  The Wilson-Fisher epsilon-expansion does not work for the three-dimensional gauge theories (there is no fixed point), and in these theories, massless in perturbation theory, there is no classical saddlepoint of finite action to use with the Lipatov\cite{lip} technique, the standard method for dealing with large-order perturbation theory.  
Even if there were such a classical saddlepoint this well-known technique would
fail in the semiclassical approximation, and the failure is precisely of the type encountered in the
Ringwald-Espinosa\cite{ring} attempt to find strong baryon violation in the EW sector\cite{rub}.  Ringwald and Espinosa were concerned with on-shell high-energy large-$N$ amplitudes in Minkowski space\footnote{This subject, and the closely-connected subject of factorial divergences in perturbation theory, engendered a great number of papers and much controversy a few years ago.  The upshot is that gauge-theory scattering processes in d=4 remain weak even at perturbative orders $N \gg 1/g^2$, but that at high temperatures such theories can become strongly-coupled.  For the author's view, see Ref. \cite{corn}; for recent summaries, see Refs. \cite{rub,vol}.}, while we are concerned with Euclidean processes, but we will see that nonetheless there is a close analogy.  In both cases the semiclassical approximation of the Lipatov technique fails to give the correct dependence on (large) external momenta.  The point in our case is that the coupling constant $g$ of the theories of interest is dimensionful ($g \sim (mass)^{1/2}$), which has the consequence that skeleton graphs for processes with large external momenta $p$ behave like $(g^2/p)^N$ in $Nth$ order perturbation theory, multiplied by a factorially-divergent factor.  Just as in the Ringwald-Espinosa problem, the semiclassical level of Lipatov (even in the massive theory, which at least has a soliton) cannot reproduce the correct dependence on external momenta, which would require quantum corrections at least to $Nth$ order to the semiclassical approach.  That is, unless {\em all} the external momenta vanish\cite{corn}, the semiclassical Lipatov approach can only yield dependence on the
external momenta which is a product of factors, one for each momentum, which is generally incorrect.  In any case, a superrenormalizable theory with only one
dimensionful parameter has no small dimensionless parameter, and is necessarily
strongly coupled, invalidating a semiclassical approach.

It is clear that d=3 gauge theories are strongly-coupled at small momentum, because they have no perturbative mass and $g^2/p$ grows large.  We know, of course, that these theories generate \cite{lind,corn82} non-perturbative masses $M \sim g^2$, and the small-momentum expansion parameter is $g^2/M = O(1)$.
It might be thought that the question of large-momentum perturbation theory is uninteresting because the $Nth$ order corrections vanish rapidly.  Unfortunately, there are
$O(N!)$ graphs, and so perturbative terms become large when $N \geq p/g^2$.
Another confounding factor for d=3 gauge theories is the lack of Borel summability, which makes it uncertain how the large-$N$ problem is to be solved by some sort of unitarization\cite{hat} or other technique going beyond perturbation theory.

We would like to study these questions for the gauge theories mentioned above, but this is too hard to do (at least for now), so we turn to a stand-in theory:
$\phi^3_5$.  This theory is, of course, irretrievably sick because its Hamiltonian is unbounded below, but that will not bother us in the present investigation.  It is related to asymptotically-free $\phi^3_6$ theory in somewhat the same way the d=3 gauge theories are related to their asymptotically-free d=4 counterparts; $\phi^3_5$ is infrared-unstable and its coupling $g$ also has dimension of $(mass)^{1/2}$.  We therefore expect to find the same phenomena discussed above for gauge theories, including the lack of Borel summability because of Borel poles on the positive real axis.

We attack the problem with direct graphical summation techniques\cite{cormor}, which have proven useful for similar problems in theories with {\em dimensionless} couplings, such as $\phi^4_4$ and $\phi^3_6$.  Specifically we will study, in $\phi^3_5$, the three-point one-particle irreducible skeleton vertex with bare internal propagators and vertices, or just the
vertex, for short. 
In any field theory, the basic graphical building blocks are the skeleton graphs for vertices and propagators, although they are little-studied at large $N$.
 These graphs have several special features, in comparison to the well-studied\cite{polk} simple ladder graphs.  As functions of the external momenta they behave as naive power-counting suggests, in the massless theory;
unlike, {\it e.g.}, ladder or planar graphs, their multiplicity is roughly
$N!$; at large $N$ the properties we will need are statistically distributed with well-defined averages and small fluctuations; and finally the bare skeleton graphs in theories with dimensionless couplings are the ones that correspond precisely to the semiclassical Lipatov approach\cite{cormor}, even at asymptotically-large momentum.

It is this last feature which is not true (at least in any way obvious to the author) in theories with dimensionful couplings, such as studied here.
While there is certainly a connection to the Lipatov approach in the massive theory, our graphical techniques correspond not just to the semiclassical first approximation, but instead to high-order corrections which instate the correct dependence on external momenta.

The connection we find to the Lipatov soliton is a subtle one, and is best illustrated by looking at the results we will derive below for the $Nth$ order contribution $\Gamma_N$ to the $\phi^3_5$ vertex, with the square of all three external momenta given by the single quantity $p^2$:  

\begin{equation}\Gamma_N \doteq N!(ag^2)^N(M^2 + \alpha p^2/N)^{-N/2}.
\end{equation}
(In this and subsequent equations the $\doteq$ symbol, to be used only at large $N$, means that we ignore non-leading terms, constant factors, and
factors of the type $N^b$ where $b$ is a fixed constant.)  Both $a$ and $\alpha$ in (1) are positive numerical constants which we will estimate.  When
$p^2 \gg M^2$ but $p^2 \ll NM^2$ it appears that the  large-$p$ momentum dependence is
of the form $\exp (-\alpha p^2/2M^2)$, but this should not be construed as characterizing the behavior of the sum over $N$ of all skeleton graphs, which also involves $N$ such that $N\gg 1$ and $p^2 \gg NM^2$.  Except for this factor, when $N$
becomes large at fixed (large) $p$, the momentum dependence goes away.
Roughly speaking, the reason for this fading away of external momentum dependence at large $N$ in the massive case is that the graphs have $O(N)$ lines  through which the momentum ``diffuses", but every one of these lines has the
mass in it.  The result is that when $N$ is so large that the momentum term in
(1) can be dropped, the sum of bare skeleton graphs beyond this value of $N$ is correctly predicted by the Lipatov technique used at zero external momentum.  From (1), the terms of this sum are $N!(ag^2/M)^N$, and one reads off the action of the Lipatov soliton as $M/ag^2$.  As we have said before, there is no such soliton (in either $\phi^3_5$ or d=3 YM) in the massless theory, where from (1) the behavior is essentially $(3N/2)!(g^2/p)^N$.

There are two ways of estimating the action of the Lipatov soliton:  One is direct numerical calculation (there are no analytic solutions), which we give below; the other is to estimate the constant $a$ from our statistical approach to skeleton graphs.  This second way disagrees with the first by about 40\%, which is comparable to the error in the graph-statistical approach as used earlier for theories with dimensionless couplings\cite{cormor}.
The graphical estimates of the constants $a,\alpha$ require a non-trivial extension of this previous work.  It involves graph-theoretic combinatorics which the author has been unable to find in the graph theory literature (as well as some known combinatorics, for which we give simple physicists' arguments).

Clearly, the next question is how to sum over $N$ in (1).  A {\em formal} Borel integral can be given for the sum, by dropping the $N!$ in (1) and doing the resultant sum to find the Borel kernel.  We will show how to express (in the $\doteq$ sense of (1)) this sum as an integral, and the formal Borel sum (which is the full skeleton vertex less the bare vertex) is a double integral of the form:
\begin{equation} \sum \Gamma_N \doteq x\int^{\infty}_0 dudt t(\frac{2e}{u})^{1/2}
e^{-t-u(y+1)}[1-(2eu)^{1/2}e^{-u}xt]^{-1}.\end{equation}
In equation (2), $x=ag^2/M$ and $y=\alpha p^2/M^2$; the Borel-transform variable of integration is $t$.  We will straightforwardly show that the singularity in $t$ associated with infrared instability occurs at $t=ag^2/M$, the inverse of the classical Lipatov soliton, independently of the external momentum.  This is, of course, what the Lipatov technique would give in the semiclassical approximation; what we get in addition by direct graphical techniques is a picture of how the summed vertex graphs depend on external momenta.  That the Borel singularity is associated with classical solitons in d=4 gauge theories is an old argument of 't Hooft\cite{tho}, which we now see is also true for lower-dimension infrared-unstable gauge theories.

Two more questions now arise:  (a) How are we to interpret these results for massive $\phi^3_5$ as having to do with d=3 gauge theories, which are massless in perturbation theory; (b) How is the Borel singularity to be regulated?
Both answers must necessarily be speculative, without independent investigation of the gauge theories; we suggest some plausible but unconfirmed answers.  As for the question of mass, we know that the d=3 gauge theories do generate a gluon mass $M \sim g^2$ non-perturbatively.  We can roughly account for this by replacing the free propagators of the bare skeleton graphs by fully-dressed propagators, using a spectral representation.
The result is that when the dressed propagators become massive, the quantity
$M$ in (1,2) is replaced by a quantity of order unity times the ``physical"
gluon mass.  So we guess that d=3 gauge theories do have graphical expansions, corrected for mass generation, somewhat like (1,2); of course, the ``expansion" parameter $g^2/M$ is just a pure number which can only be calculated non-perturbatively.

As for regulating Borel singularities, we speculate that this is done much the same way as is known to happen\cite{bog} in other models, where the semiclassical Lipatov solution (actually a sphaleron-like solution for gauge theories) conspires with the Borel sum of perturbation theory to remove unphysical imaginary parts.  This amounts to replacing $e^{-t}$ in (2) by
$e^{-t}-e^{-t_0(u,x)}$ where $t_0(u,x)$ is the position of the pole in $t$ in equation (2).  We will discuss this further in Sec. 5.  The result is that the regulated sum corresponding to (2) is strongly-coupled because $ag^2/M$ is  $O(1)$, but strong coupling has nothing to do with the external momenta.  This is because the mass regulates the infrared singularities of the massless theory, and the Lipatov soliton regulates the problems mentioned above which could occur when $p\gg g^2$ but $p\leq Ng^2$.

In the regulated sum corresponding to (2) it is straightforward to show that the leading asymptotic behavior at large $p$ is $\sim g^2/p$, just as one would expect from one-loop corrections and the naive argument that higher-order graphs are suppressed by more powers of $p^{-1}$.

Although there is no obvious way to associate (1) or (2) directly with a semiclassical Lipatov analysis, one gets the feeling that there really is something semiclassical behind the direct graphical approach.  This is certainly so\cite{cormor} for theories with dimensionless couplings, where the skeleton graphs so far analyzed are essentially independent of momentum even when these are large, and therefore are in effect zero-momentum graphs, where the Lipatov technique surely works.  And even for our theories with dimensionful couplings, the graphical analysis with its emphasis on well-defined average properties of a very large number of graphs, and small fluctuations around these averages, suggests some underlying mean-field theory for their descriptions.  This theory is presently unknown, and it would be very helpful to find it. 

\section{Nth Order Bare Skeleton Vertex in $\phi^3_5$}

The Euclidean action is:

\begin{equation}  S = \int d^5x [ \frac{1}{2} \partial_i \phi(x)^2 
+ M^2\phi(x)^2 + \frac{g}{3!}\phi(x)^3 ].
\end{equation}
We will be concerned for the most part with the behavior of the bare skeleton  vertex function, which is dimensionless when an overall power of $g$ is removed. The vertex at $Nth$ order of perturbation
theory, which we denote $\Gamma_N$, involves $g^{2N}$ divided by the $Nth$ power of some quantity with the
dimensions of mass.  This quantity will certainly be the mass (if there is one)
when all external momenta vanish, but it may involve external momenta too, when these are large compared to $M$.  Exactly what happens depends on the graphs considered.  Vertex graphs which are two-particle-reducible, such as the graph of Fig. 1, in general depend on both.  For example, let $p_1$ vanish, and $p_2^2 = p_3^2 \equiv p^2 \gg M^2$.  One simply approximates the propagator joining
the vertices carrying momentum $p$ by $1/p^2$ to find that this graph is the product of this factor and a self-energy graph at zero momentum, which can only depend on $g^2/M$.  More elaborate versions of this sort of behavior were studied long ago\cite{polk}.  All of these older studies of graphs emphasize infrared singularities arising from highly-reducible graphs, such as sums of ladder graphs, but there is only one such (bare) ladder graph at every order $N$, so we cannot hope to study factorial behavior there.

Bare skeleton graphs, however, have factorial multiplicity, and should correspond to the perturbative terms found by the Lipatov analysis.  But at the semiclassical level this analysis decouples almost completely from the external momenta, as we will discuss in Sec. 4.  To find the momentum dependence even approximately requires a graphical analysis.

Consider an $Nth$-order vertex skeleton graph with bare vertices and propagators, an example of which is shown in Fig. 2.  Such a graph, in graph-theoretic language, has girth four, which means that no loop in it has less than four lines; it has $N$ loops, $3N$ lines, and $2N+1$ vertices.   It is both one- and two-particle-irreducible (except trivially by cutting the lines incident on any of the three external vertices).  In $\phi^3_5$, any graph has the value, expressed in terms of Feynman-parameter integrals after the momentum integrals have been done,
\begin{equation} \Gamma(G)= S(G)\Gamma(N/2)[\frac{g^2}{(4\pi )^{5/2}}]^N
\int [dx]_{3N}U^{-5/2}(\phi/U+\sum x_iM^2_i)^{-N/2}. \end{equation}
Here
\begin{equation} [dx]_{3N}= dx_1\dots dx_{3N}\delta (1-\sum x_i),
\end{equation}
$U$ is the determinant of the graph, a sum of positive $Nth$-order monomials in the Feynman parameters $x_i$ constructed according to rules reviewed briefly below, and $\phi$ is a sum of positive monomials of order $N+1$ in the
$x_i$, multiplied by one or another of the $p_i^2$ according to rules reviewed below.  The $M_i^2$ are the squared masses of the internal lines; for later applications we distinguish them, but in the bare skeleton graphs each mass is just the mass $M$.  Finally, $S(G)$ is the symmetry number of the graph; at large $N$, almost all (in the $\doteq $ sense) graphs have symmetry number of unity, and we will ignore this factor in what follows.

The non-trivial difference between what we do here and previous work
\cite{cormor} is the appearance of the $\phi/U$ term in (4) to a power other than zero.
For a dimensionless skeleton vertex in a theory with dimensionless couplings, this term either appears to the zeroth power or as a single logarithm (which can be eliminated by taking the Minkowski-space imaginary part).  In such cases there is no dependence on external momenta at all, and these might as well be zero.  This means that the sum of (the imaginary part of) skeleton graphs at  large order must reproduce the semiclassical Lipatov value, which is only  completely correct when all the momenta vanish.  But in the present case the
$\phi/U$ term is present and must be dealt with.  This is, by the way, exactly what must be done to understand the perturbation-theory analog of the Ringwald-Espinosa problem, which in part stems from the fact that graphs with a large ($O(1/\lambda )$) number of external legs carrying momentum, in a theory with dimensionless coupling $\lambda$,  have an equally-large negative power of some combination of the external momenta.  This negative power comes from the
$\phi/U$ term. The semiclassical Lipatov approach does not give this dependence correctly.

We will deal with this $\phi/U$ term just as was done earlier\cite{cormor} for the case when it did not appear:  Just replace $\phi$ as well as $U$ in (4) by their average value (averaged over the Feynman-parameter integral).  Such an approach certainly cannot work for many types of $Nth$-order graphs, for example, the vertex ladder graph.  Ladder and other highly-correlated graphical structures have $\phi$-functions which can vanish very rapidly compared to
$U$, and thereby generate power-law infrared singularities (that is, inverse powers of external momenta which would naively be expected when the momenta are large get replaced by inverse powers of the mass $M$; this is how Regge poles are generated\cite{polk} in $\phi^3_4$).  But large-$N$ skeleton graphs are very random, and one cannot make $\phi$ vanish without making $O(N)$ Feynman parameters small\footnote{When all the parameters of $J$ loops are made small, $U$ vanishes like the $Jth$ power of the smallness parameter, so there are cancellations in the ratio $\phi/U$.}; the region of parameter space where this can happen is too small to lead to an infrared singularity.  This statement is just another way of stating Weinberg's\cite{wein} theorem.  So we feel justified to replace both 
$\phi$ and $U$ by their parameter-space averages, as was earlier justified simply for $U$ alone\cite{cormor}.  

The rules for $U$ and $\phi$ are well-known\cite{nak}. To find $U$, find the {\em chord sets}, which are the sets of all lines which, if cut, turn the graph into one connected tree, called a spanning tree..  The monomials of $U$ are the products of $x_i$ constituting the chord sets.  To find $\phi$, find the {\em cut sets}, or lines which, if cut, turn the graph into exactly two connected trees.  Write the product of $x_i$ for each cut set, multiply the result by the square of the sum of the external momenta flowing into or out of either tree, and sum; the result is $\phi$.

As discussed above, we argue that for fixed and large order $N$ the sum of all bare skeleton graphs of that order is (in the $\doteq $ sense) reasonably well-estimated by replacing $U$ and $\phi$ in equation (4) by their averages over the Feynman-parameter integral, and multiplying (4) by $Q_N$, the number of graphs of $Nth$ order.  It is easy to compute the parameter-integral average of any monomial; the two of concern are $O(N)$ and $O(N+1)$ and we have
\begin{equation} \langle x_1\dots x_K \rangle = \frac{\Gamma(3N)}{\Gamma(3N+K)} \end{equation}
so
\begin{equation} \langle x_1\dots x_N \rangle \doteq \frac{(27/256)^N}{N!};
\end{equation}
\begin{equation} \langle x_1 \dots x_{N+1} \rangle = \frac{1}{4N}
\langle x_1\dots x_N\rangle. \end{equation}

It follows that $\langle U \rangle $ is found by multiplying the result in (7) by the {\it complexity} $C_N$ of the graph, that is, by the number of spanning trees.
This number has been found\cite{cormor} numerically earlier:
\begin{equation} C_N \doteq (\frac{16}{3})^N.
\end{equation}             
Then
\begin{equation} \langle U \rangle = C_N \langle x_i\dots x_N \rangle \doteq 
(\frac{9}{16})^N\frac{1}{N!} \end{equation}.

The estimation of $\phi$ is a little more complicated.  First, we need the number of cut sets of the graph, and in the Appendix we show that this number is
just $2C_N$.  However, this is not exactly what we need.  Each cut set monomial is multiplied by the square of the momentum entering either of the two trees;
sometimes this is one of the external momenta $p_i^2$ and sometimes it is zero
(see Fig. 3).  In the Appendix we give a sketchy argument that the fraction $\beta$ of cut sets containing an external momentum is 1/2.  Since each monomial in
$\phi$ has the same average, it is clear that $\langle \phi \rangle $ is the same if we replace each $p_i^2$ by the average squared momentum $p^2$:
\begin{equation} p^2 \equiv \frac{1}{3}(p_1^2+p_2^2+p_3^2). \end{equation}
Then using these remarks and equation (8) we find:
\begin{equation} \langle \phi \rangle = 2\beta \frac{p^2}{4N} \langle 
U \rangle = \frac{p^2}{4N} \langle U \rangle \end{equation}.

Finally we need the elementary integral
\begin{equation} \int [dx]_{3N}=\frac{1}{\Gamma(3N)}\end{equation}
and the number $Q_N$ of $Nth$-order $\phi^3$ graphs.  This number is known\cite{bc}, and we give a physicists' derivation of it in the Appendix:
\begin{equation} Q_N \doteq N!(\frac{3}{2})^N.  \end{equation}
This number includes all graphs, skeleton or otherwise, but as the derivation in the Appendix shows, almost all the graphs are skeleton graphs.  

We now replace $\phi$ and $U$ in (4) by their averages, and multiply by
$Q_N$ to find an estimate for the sum of $Nth$-order skeleton graphs as given in equation (1), with
\begin{equation} a= \frac{2^{17/2}}{3^7(4\pi )^{5/2}}= 2.96 \times 10^{-4},\;\;\alpha = \frac{1}{4}
\end{equation}.

Three remarks:  First, at $p_i=0$ the graphical expansion should yield the Lipatov results.  We study this in Sec. 4, finding for the coefficient $a$ the value $4.55 \times 10^{-4}$.  The graphical value is about 2/3 the Lipatov value, and is less than it, in line with general arguments\cite{cormor} based on H\"older inequalities.  Second,
one might compare the above value for $a$ with that found by graphical techniques for $\phi^3_6$, where $a$ is defined by the expansion
\begin{equation} \Gamma_N \doteq N!(ag_6^2)^N \end{equation}
and the dimensionless coupling $g_6$ appears in the action just as it does in 
equation (3).  Define the dimension as $d=6-\epsilon$; then the present
five-dimensional results and the earlier\cite{cormor} six-dimensional results can be written in one formula as:
\begin{equation} a= \frac{2^{11-5\epsilon /2}}{3^{8-\epsilon /2}
(4\pi )^{3-\epsilon/2}}. \end{equation}
This is a kind of $\epsilon$-expansion which can be used for infrared-unstable
theories.  Third, when one goes beyond bare skeleton graphs by dressing the vertices and propagators in such a graph, one will find logarithmic terms 
which presumably will change powers of momenta and lead to anomalous dimensions.
This is a subject for future analysis.

\section{The Formal Borel Sum}
We have now established the result (1) (repeated below for convenience)

\begin{equation}\Gamma_N \doteq N!(ag^2)^N(M^2 + \alpha p^2/N)^{-N/2}
\end{equation}
and would like to know how to write it in the usual Borel-integral form.  Of course, this integral will have a singularity on the positive real axis, because all the terms of (18) are positive.  We will discuss regulation of this singularity later.

Given a formal sum
\begin{equation} h(x) \equiv \sum c_N x^N,  \end{equation}
the Borel prescription is
\begin{equation} h(x) = \int_0^{\infty} dt e^{-t} g(xt) \end{equation}
where  
\begin{equation} g(x) = \sum \frac{c_N}{N!}x^N.  \end{equation}
So from (18) we need the sum
\begin{equation} g(x,y) \equiv \sum_1 x^N(1+y/N)^{-N/2};\;\;x=ag^2/M,\;
y=\alpha p^2/M^2. \end{equation} 
This non-standard sum can be written (in the $\doteq $ sense) as an integral, using the identity
\begin{equation} (y+N)^{-N/2} = \frac{1}{\Gamma(N/2)}\int_0^{\infty}
du e^{-u(y+N)}u^{(N/2)-1}. \end{equation}
Then Stirling's formula gives
\begin{equation} (1+y/N)^{-N/2} \doteq (2e)^{N/2}\int_0^{\infty}
du e^{-u(y+N)}u^{(N/2)-1}. \end{equation}
Use this to do the sum in (22):
\begin{equation} g(x,y) \doteq x\int_0^{\infty} due^{-u(y+1)}(\frac{2e}{u})^{1/2}[1-x(2eu)^{1/2}e^{-u}],  \end{equation}
and from (20) the formal Borel sum of (18) is equation (2), repeated for convenience:
\begin{equation} \sum \Gamma_N \doteq x\int^{\infty}_0 dudt t(\frac{2e}{u})^{1/2}
e^{-t-u(y+1)}[1-(2eu)^{1/2}e^{-u}xt]^{-1}.\end{equation}

The singularities of this double integral are endpoint branch cuts, expressing a dependence on $p$ rather than $p^2$, and a Borel pole where the denominator in
(26) has a second-order zero, or pinch singularity.  (A single zero not at an endpoint can be evaded by shifting the contour.)  The second-order zero occurs where the function $(2eu)^{1/2}e^{-u}$ has a maximum, which happens at
$u=1/2$, and the maximum value is unity.  Therefore, the Borel singularity is at
$t_0=1/x=M/ag^2$, and the Borel ambiguity is $e^{-t_0}=e^{-M/ag^2}$.  This is, as expected, the exponential of a classical soliton action associated with the
Lipatov technique.  So we should be able to find the same action ({\it i.e.}, the same value of $a$) by looking for classical solitons of $\phi^3_5$.  This we do in the next section.    

\section{Lipatov Analysis and the Need for Mass}

We briefly review the Lipatov technique, showing how it fails at the semiclassical level to get the right dependence on external momenta.  Then we show that the technique fails, even at zero momentum, for zero mass, and finally we find the appropriate soliton for non-zero mass.

Write the $Nth$ order contribution to the  full (reducible) vertex in path-integral form:
\begin{equation}  Z^{-1} \int (d\phi ) e^{-S_0}[-\frac{g}{3!}
\int d^5x \phi(x)^3]^N \prod_i \int d^5x \exp (ip_i \cdot x_i) \phi(x_i)
\end{equation}
where $S_0$ is the free action.  At large $N$ the external wave-function factors are to be evaluated by inserting the classical solution $\phi_c(x_i-a)$ and integrating over the translational collective coordinate $a$; the result is a product of three factors each depending on only one external momentum.  This is clearly wrong for the skeleton graph of Fig. 2, unless all the momenta vanish.
But if all the momenta vanish a mass is necessary to construct a dimensionless expansion parameter $g^2/M$.  The smaller the mass the more strongly coupled
the theory is, but in an infrared-unstable theory massless at the Lagrangian level (including $\phi^3_5$) there is a lower bound on the mass, arising from tachyonic poles in the zero-mass limit of such theories\cite{corn82,corn96}.

The Lipatov technique understands this non-existence of the massless theory
at zero momentum as the non-existence of a massless soliton with finite action (of course, the massless soliton is precisely what does arise in theories with dimensionless couplings).  The classical equations coming from the saddlepoint of the action\footnote{We can now drop the external wave-function factors.}
in equation (27) above is:
\begin{equation} \Box \phi - M^2 \Phi +\frac{3N\phi^2}{K} = 0;\;\;
K=\int d^5x \phi^3.
\end{equation}
Let us look for a spherically-symmetric solution for the massless case.
Even though the underlying field theory has a dimensionful coupling constant, it is absent in (28) and this equation is independent of the scale of length chosen (there is a collective coordinate for scale).  Call this scale $\rho$ and change variables to:
\begin{equation} \phi(x) = \frac{K}{3Nx^2}q(x^2/\rho^2);\;\;\frac{x^2}{\rho^2}
\equiv e^{-t}.  \end{equation}
(Note that from (28) $K \sim \rho^{1/2}$ so $\phi$ has its canonical mass dimension of 3/2.)  Performing this change of variables in (28) yields
\begin{equation} \ddot{q} + \frac{1}{2} \dot{q} + \frac{1}{4}q^2 - \frac{1}{2}
q = 0. \end{equation}
The dots indicate derivatives with respect to $t$, so this is the equation of motion of a particle of coordinate $q$ with friction in a potential
\begin{equation} V(q) = \frac{q^3}{12} - \frac{q^2}{4}. \end{equation}
This potential has a relative maximum at $q=0$ and a minimum at $q=2$; it reaches the value zero at $q=0,3$ so that without friction there would be a soliton going between these two points.  But this is impossible with friction, and any solution vanishing at $x=\infty\;\;(t=-\infty )$ necessarily settles to the minimum at $q=2$. Indeed, $q=2$ or from (29) $\phi \sim 1/x^2$ is an exact solution, but with infinite action.  

There is an analog of this in d=3 YM theory.  The massless version has only one spherically-symmetric soliton, the Wu-Yang soliton, and it has infinite action too.  Indeed, one can repeat the steps of writing down the equations and changing variables rather as in (28, 29) to find the same sort of motion-with-friction analog.  So the Lipatov analysis fails here, too, as we know it must.

Now turn to the massive case, and write
\begin{equation} \phi(x) = (\frac{KM^2}{3N})p(u);\;\;u \equiv M^2x^2.\end{equation}  
The equation for $p(u)$ is:
\begin{equation}  p'' + \frac{5}{2}\frac{p'}{u} + \frac{1}{4u}(p^2-p)=0
\end{equation}and it does have a finite-action solution (with $\phi$ approaching a constant at $x=0$ and vanishing exponentially at infinity).  We have determined the solution and the action numerically.  One can then easily evaluate the coordinate integrals in (27) and thereby find the semiclassical Lipatov-technique value for the contribution $\Gamma_N$ to the $Nth$-order vertex at zero momentum:
\begin{equation} \Gamma_N \doteq N!(\frac{ag^2}{M})^N;\;\;a=4.55 \times
10^{-4}.  \end{equation}
This value of $a$ is to be compared to the graph-technique value of
$a=2.96 \times 10^{-4}$ given in (15).

Again, the existence of a massive soliton has its analog in d=3 YM theory, which also has one\cite{corn77}.  Its action is $4\pi \times 5.4M/g^2$.  Of course, the mass does not appear in the clasical action, but is generated by non-perturbative quantum effects\cite{lind,corn82} which can be described in terms of an effective massive action, which does have a soliton.  Even thought $M\sim g^2$, the mass vanished to all orders of perturbation theory.  So although it is essential to include the mass in order to stabilize the theory, once one does so in, {\it e.g.}, the bare skeleton graphs one has really gone beyond perturbation theory altogether.  In the next section we speculate on this and other non-perturbative features.

\section{Beyond Perturbation Theory}

Aside from the question of how to incorporate a non-perturbative mass in skeleton-graph expansions like (1) or (2), there is the problem of regulating the Borel singularity in (2).

In $\phi^3_5$, if not in d=3 gauge theory, it is straightforward to incorporate dressed propagators into the bare skeleton graphs.  One writes for the propagator on line $i$ a Lehmann representation:
\begin{equation} \Delta (p_i) = \int dM_i^2 \frac{\rho (M_i^2)}{p_i^2
+M_i^2};\;\;\int dM_i^2 \rho (M_i^2) = 1 \end{equation} so that the right-hand side of equation (4), which gives the contribution of the $Nth$-order graph $G$, becomes  

\begin{equation} \Gamma(N/2)[\frac{g^2}{(4\pi )^{5/2}}]^N \prod \int dM_i^2 \rho (M_i^2)
\int [dx]_{3N}U^{-5/2}(\phi/U+\sum x_iM^2_i)^{-N/2}. \end{equation}
As before we will replace functions of the Feynman $x_i$ by their averages, and
$\langle x_i \rangle = 1/3N$, so the $\sum x_i M_i^2$ term in the denominator
is replaced by the average value $\sigma$:
\begin{equation} \sigma \equiv \frac{1}{3N}(\sum M_i^2).  \end{equation}
To the extent that the positive weight functions $\rho(M_i^2)$ behave like probability distributions we can invoke the central limit theorem to replace $\sigma$ by the average of $M_i^2$ over $\rho$, clearly a quantity differing from the physical mass
$M^2$ by a factor of order unity.  So our main results (1,2) continue to hold, with $M^2$ defined as this average.  

It is somewhat delicate to apply this line of argument to d=3 gauge theory, for several reasons.  First, one cannot maintain gauge invariance without dressing vertices as well as propagators; second, even if the internal vertices are dressed in the skeleton graphs, the final result is gauge-dependent and must receive corrections such as used in the pinch technique\cite{corn82,cornpa}; third, in any theory where there is no mass at the classical level, the quantum mass must depend on the propagator momentum $p^2$ and vanish at large $p^2$.
It is known\cite{lav} that this running mass vanishes like $g^2\langle G_{ij}^2 \rangle p^{-2}$.  However, presumably one should not replace $M^2$ in (1) or (2) by $M^2(p^2)$, but rather by $M^2(p^2/N)$, just as the graphical analysis tells us to do for the kinetic-energy terms of propagators.  

Arguments like this aside, there is another reason to believe that it may be a decent approximation for d=3 gauge theory to use (1,2) with the zero-momentum quantum mass
$M$ in it, thus going beyond perturbation theory.  This is that the Lipatov-technique result at zero momentum is likely to be the right starting place to understand large-order dressed skeleton graphs, and this technique will incorporate whatever quantum mass is generated in the full theory.

Now turn to the problem of regulating the Borel singularity.  It is known\cite{bog} in various quantum-mechanical and field-theoretic problems that the Borel singularity, which may give rise to an imaginary part having no physical interpretation, is cancelled by the contribution of solitons such as the one we have found.  We do not attempt to study this problem in detail here, but simply suggest a plausible answer, which is to subtract off an exponential term in the Borel integral which gets rid of the singularity.  So (2) or (26) gets modified to 
\begin{equation} \sum \Gamma_N \doteq \int^{\infty}_0 dudt xt(\frac{2e}{u})^{1/2}
e^{-u(y+1)}[e^{-t}-e^{-t_0(x,u)}][1-(2eu)^{1/2}e^{-u}xt]^{-1}.\end{equation}
where
\begin{equation} t_0(x,u)=\frac{e^u}{x(2eu)^{1/2}}.  \end{equation}
It can now be easily checked that there is no sign of a large contribution from large momentum $p$, as one might have naively expected from a behavior like
$N!(g^2/p)^N$ when $N$ is large and $p \sim Ng^2$.  The leading large-$p$ behavior is what is expected from one-loop corrections, namely $g^2/p$.

\section{Conclusions}

In this paper we have studied the large-order perturbative behavior in theories where external momenta are expected to play a significant role.  The two cases of interest are in theories where the coupling constant is dimensionful (such as d=3 gauge theory or $\phi^3_5$) or, in d=4 gauge theories, the case of multiparticle amplitudes where the number of particles is $O(1/\alpha)$\cite{ring,rub,corn,vol}.   The troublesome point is that the widely-used Lipatov technique cannot sense these external momenta in the semiclassical approximation.  To see their effect one has to do something much like the graphical analysis we give here, or, much better, find a Lipatov technique based on some sort of effective field theory which allows incorporation of the external momenta. 

Even at zero momentum in d=3 gauge theory, one must use solitons of an effective action incorporating a non-perturbative gluon mass.  Otherwise the Lipatov technique falls apart, for lack of a massless soliton with finite action.
Strictly speaking perturbation theory at large order never makes sense in this theory, because we have found that both small-$p$ and large-$p$ problems (the latter coming from large $N$) are resolved by a finite mass, which is absent in perturbation theory.  In massless $\phi^3_5$ theory at large $p$ one can always go to $N$ so large (of order
$(p/g^2)^{2/3}$, based on (1)) that perturbation theory breaks down, quite the opposite of what one expects in a superrenormalizable theory, where higher orders are supposed to fall off more rapidly in $p$.  However, in the massive theory as $N$ increases the momentum dependence is overtaken by a mass dependence such that the expansion parameter becomes $g^2/M$.  Perturbation theory still breaks down, but because of the infrared singularities of the theory.    

It would, of course, be valuable to extend our graphical work to such a gauge theory, but this seems to be very hard indeed at present.  Again, an effective-action approach would be much better.   

The graphical analysis goes beyond earlier large-$N$ work in including the effect of dimensionful denominators through the $\phi$ function of Sec. 2.
It would be interesting to extend this work to problem of multiparticle amplitudes in d=4 renormalizable theories, to see if the kind of random-graph arguments we use here and earlier\cite{cormor} can be made to work.  What we have done is a minimum first step. In view of the mass dimensionality of d=4 scattering amplitudes, which is $4-N$ for an $N$-point vertex, a behavior like $(\lambda/p)^N$ in a theory with {\em dimensionless} coupling $\lambda$ corresponds to only that part of the high-energy fixed-angle amplitude in which the order of perturbation theory is correlated with the number of legs.  More work is needed to study the general case.

\newpage
\begin{center}
{\Large \bf Acknowledgements}
\end{center}
This work was supported in part by the National Science Foundation under grant
PHY 9531023.

\newpage
\appendix
\begin{center}
{\Large \bf Appendix:  Graphical Combinatorics}
\end{center}
Here we discuss three questions:  (a) The number of $Nth$-order $\phi^3$ graphs;
(b) The number of cut sets at this order; (c) The fraction of cut sets which carry finite momentum.  

To count graphs, we can construct an $N+1st$-order graph by adding one line to
any $Nth$-order graph in all possible ways.  There are $(3N+3)^2/2$ ways to do this (we never consider graphs with one-line loops, and the added line can begin or end on any of the $3N$ internal lines or the three external lines).  But this overcounts by a factor of $3N+3$.  To see this, mark any other line in the $Nth$-order graph (not the one added to make the $N+1st$ order).  Exchange the marked line and the
added line.  This gives the same graph in $N+1st$ order, but constructed from a different graph at $Nth$ order.  So the relation between the number of graphs
$Q_{N+1}$ at $N+1st$ order to $Q_N$ is
\begin{equation} Q_{N+1} = \frac{3}{2}(N+1)Q_N;\;\; Q_N \doteq
(\frac{3}{2})^N N!. \end{equation}

Almost all of these graphs are skeleton graphs.  Let the $Nth$-order graph be a skeleton.  Then, having chosen one of $3N+3$ positions to place one end of the added line, there are only $3N$ ways to place the other (it cannot go on the line it started, or on lines separated from it by only one vertex).  Continuing as above, one finds that $Q_N$ for skeleton graphs differs from that for all graphs by a factor $N^c$, where $c$ is a fixed constant.  So in the $\doteq $ sense the two counts are the same.  For a full discussion of the graph-counting problem, see ref. \cite{bc}.

Next we count the cut sets, which enter the definition of $\phi$.  A good heuristic for this is to start from the complexity, or number of spanning trees.
The complexity $C_N$ is easily bounded:  To form a tree, one must cut $N$ out of
$3N$ lines, so
\begin{equation} C_N \leq \frac{(3N)!}{(2N)!N!} \doteq (\frac{27}{4})^N
\end{equation}.
The actual number, computed numerically\cite{cormor}, is $(16/3)^N$.  The $Nth$ roots differ by only 25\%.  The true number is smaller than $27/4$ because not every set of $N$ cuts yields a connected graph with no loops.

We can find the cut sets by cutting the tree one more time; that is, a cut set comes from the original graph by making $N+1$ distinct cuts.  Thus the number of cut sets $\tilde{C}_N$ is bounded by:
\begin{equation} \tilde{C}_N \leq \frac{(3N)!}{(N+1)!(2N-1)!} \doteq 
2(\frac{27}{4})^N.
\end{equation}
This suggests that $\tilde{C}_N = 2C_N$, for which we now argue in the same way we found $Q_N$.  Take a spanning tree and mark any line not already cut;
there are $2N$ ways to do this.  Now exchange the marked line with any line already cut.  Any of the $N$ ways of doing this exchange repeats a graph, just as we argued for $Q_N$.  Therefore we divide $2N$ by $N$ to come to
$\tilde{C}_N \doteq 2C_N$.

Not in all cases does one of the two trees have external momentum flowing into (or out of) the tree (recall Fig. 3). We estimate this fraction from a particular set of graphs of the type in Fig. 4, in which $N$ vertices lie on the top line and another $N$ on the bottom; these are joined in all possible ways by lines which we call crossing lines.  There are clearly $N!$ such graphs, less than $Q_N$, but we have no reason to believe our argument fails for more complicated graphs.  

First we make a spanning tree by cutting the line 1-2 a certain number of times, the line 1-3 a certain number of times, and the crossing lines a certain number of times; an example is shown in Fig. 5, in which two of the crossing lines remain uncut, and there are J cut crossing lines between vertices 1 and A, as well as K cut crossing lines between B and 3.  We need to cut one more line to make a cut set.  Evidently, if we cut any of the J (noncrossing) lines between vertices 1 and A, or any of the K lines between
B and 3, we form a graph of two connected trees both of which have a total of zero external momentum flowing into the graph; we call this a null cut.  Any other cut leads to trees carrying one of the external momenta $p_i$; we call this a non-null cut.  Elementary combinatorics which is not worth recording here shows that, on the average when $N$, J, and K are all large,  the ratio of null cuts to non-null cuts is just L/(L+2), where L is the number of uncut crossing lines.  This ratio approaches unity as L becomes large, so we conclude that about half the cuts are non-null and enter into the expression for $\phi$.
To form $\phi$ all the Feynman parameters in a non-null cut are multiplied together and then multiplied by the square of the external momentum flowing into  either tree.  As explained in the text we replace $\phi$ by its average wherever it appears.  Since the average of every monomial of $N+1$ Feynman parameters has the same value, it is evident that $\langle \phi \rangle $ is a symmetric function of the 
$p_i^2$, and can be calculated by multiplying the monomial average by the number of non-null cut sets and by the average squared momentum $p^2$:
\begin{equation} p^2 = \frac{1}{3}(p_1^2+p_2^2+p_3^2). \end{equation}        
\newpage
\begin{center}
{\Large \bf Figure Captions}
\end{center}
Fig. 1.  A two-particle-reducible graph.\\
Fig. 2.  A bare skeleton graph:  No internal propagator or vertex parts.\\
Fig. 3.  A graph cut into two connected trees.  Neither tree carries net external momentum flowing into the graph.\\
Fig. 4.  A skeleton graph drawn by connecting $N$ vertices on line 1-2 with $N$ vertices on line 1-3.\\
Fig. 5.  A spanning tree for a graph of the type in Fig. 4.
 
\newpage
\epsfig{file=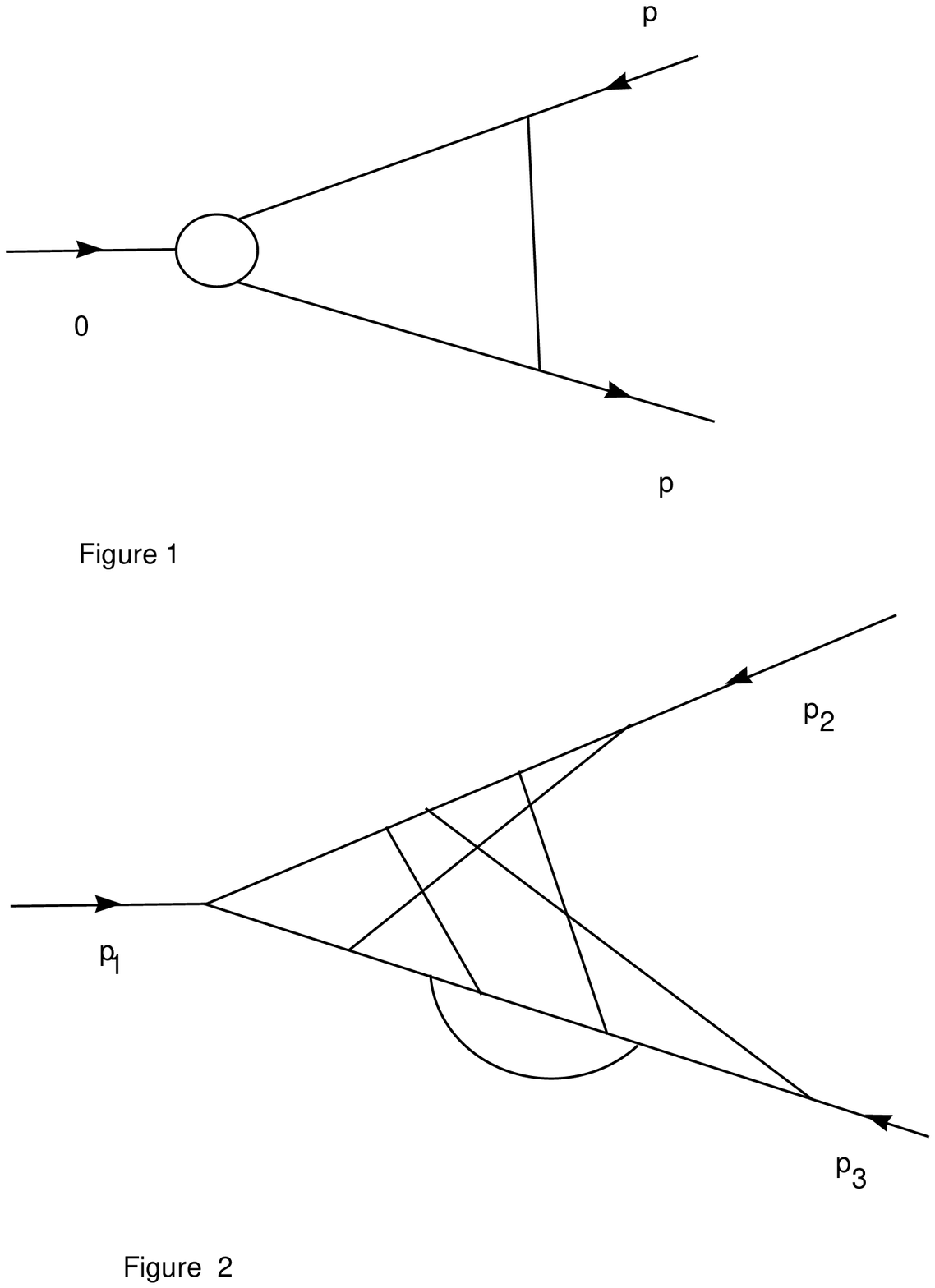,height=7in,clip=}
\newpage
\epsfig{file=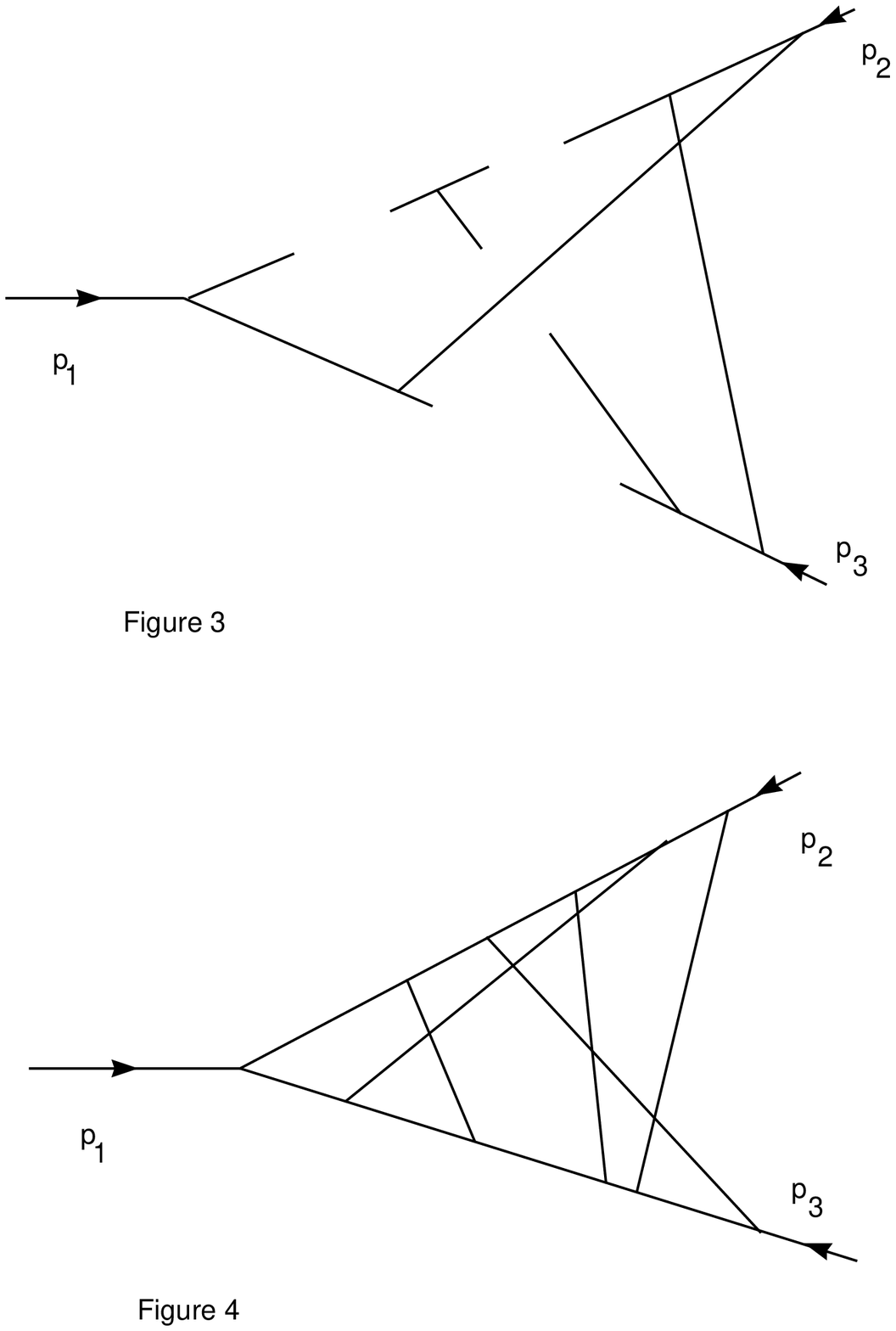,height=7in,clip=}
\newpage
\epsfig{file=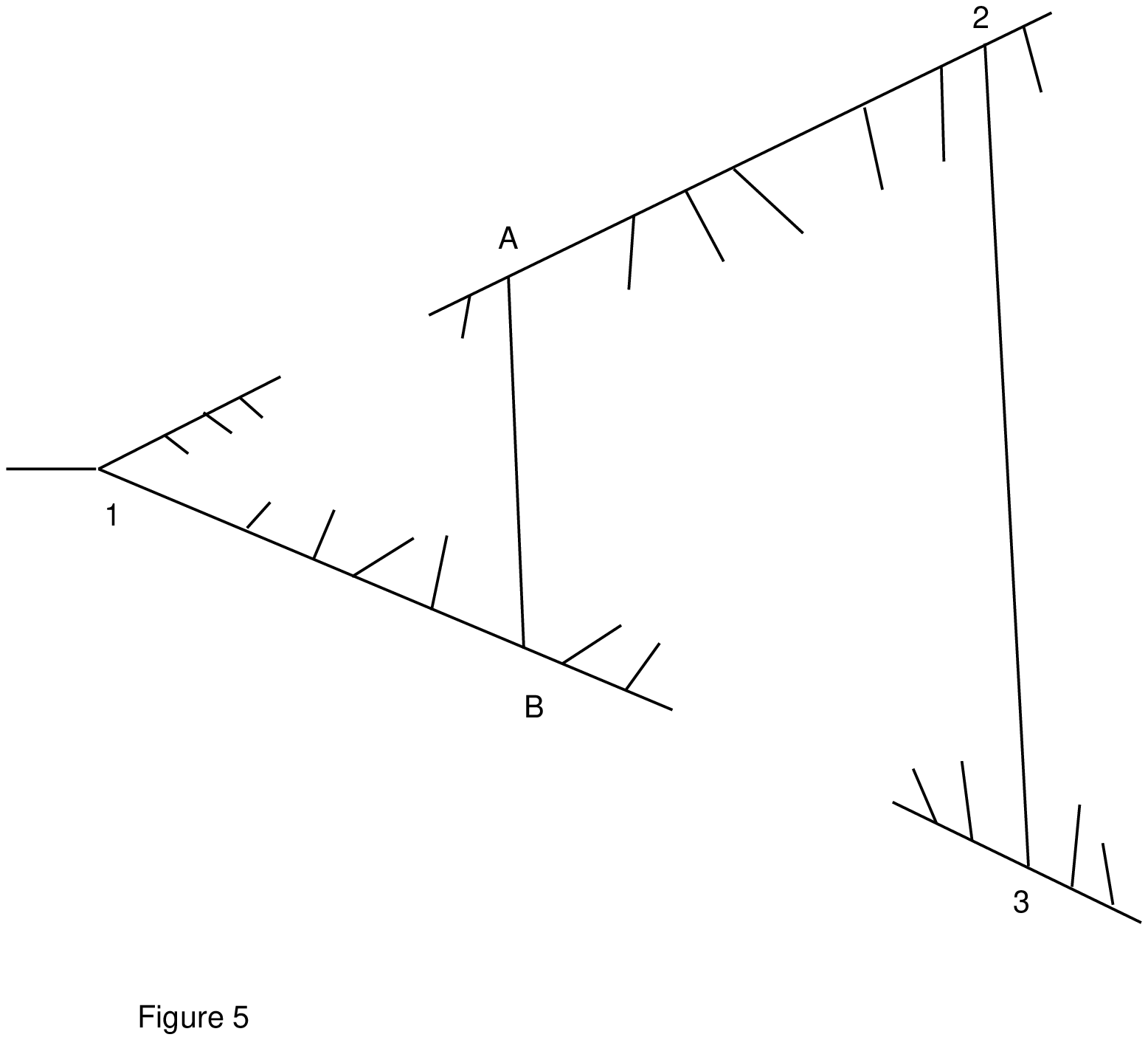,height=5in,clip=} 

\newpage
\begin{thebibliography}{99}
\bibitem{lip}  L. N. Lipatov, Pisma Zh. Eksp. Teor. Fiz. {\bf 14}, 179 (1976);
E. Br\'ezin, J. C. le Guillou, and J. Zinn-Justin, Phys Rev. D {\bf 15}, 1544, 1588 (1977).  A collection of reprints and further references is {\it Large-Order Behavior of Perturbation Theory}, edited by J. C. le Guillou and J. Zinn-Justin, Currrent Physics Sources and Comments Vol. 7 (North-Holland, Amsterdam, 1990).  
\bibitem{ring} A. Ringwald, Nucl. Phys. {\bf B330}, 1 (1990); O. Espinosa,
Nucl. Phys. {\bf B343}, 310 (1990).
\bibitem{rub} There have been hundreds of papers written on baryon/lepton violation with emphasis on Ringwald-Espinosa-like processes; for a comprehensive review, see V. A. Rubakov and M. E.Shaposhnikov, ``Electroweak Baryon Number Nonconservation in the Early Universe and in High-Energy Collisions", CERN preprint CERN-TH-96-13, hep-ph/9603208 (unpublished).
\bibitem{corn} J. M. Cornwall, Phys. Lett. B{\bf 243}, 271 (1990); J. M. Cornwall and G. Tiktopoulos, Phys. Rev. D {\bf 45}, 2105 (1992), D {\bf 47}, 1629 (1993).
\bibitem{vol} M. B. Voloshin, talk at the 27th International Conference on High-Energy Physics, Glasgow, Scotland, July 1994; ICHEP 1994 Proceedings, p. 121.
\bibitem{lind} A. D. Linde, Phys. Lett. {\bf 96B}, 289 (1980).
\bibitem{corn82} J. M. Cornwall, Phys. Rev. D {\bf 26}, 1453 (1982); J. M. Cornwall, W.-S. Hou, and J. E. King. Phys. Lett. {\bf 153B}, 173 (1985); J. M.
Cornwall and B. Yan, Phys. Rev. D {\bf 53}, 4638 (1996).
\bibitem{hat} N. Hatzigeorgiu and J. M. Cornwall, Phys. Lett. {\bf B327}, 313 (1994). 
\bibitem{cormor} J. M. Cornwall D. A. Morris, Phys. Rev. D {\bf 51}, 4844 (1995);
D {\bf 52}, 6074 (1995).
\bibitem{polk} J. C. Polkinghorne, J. Math. Phys. {\bf 4}, 503 (1963).  Polkinghorne works with $\phi^3_4$ and encounters logarithmic, not power, dependence on external momenta.
\bibitem{tho} G. 't Hooft, in {\it Deeper Pathways in High-Energy Physics}, edited by A. Perlmutter and L. F. Scott (Plenum, New York, l977), p. 699.
\bibitem{bog}  E. Bogomol'nyi, Phys. Lett. {\bf 91B}, 431 (1980); J. Zinn-Justin, Nucl. Phys. {\bf B192},125 (1981); {\it Quantum Field Theory and
Critical Phenomena} (Oxford, New York, 1996).
\bibitem{coll} A. J. MacFarlane and G. Woo, Nucl. Phys. {\bf B77}, 91 (1974);
J. C. Collins, {\it Renormalization} (Cambridge University Press, Cambridge, England, 1984).
\bibitem{wein} S. Weinberg, Phys. Rev. {\bf 118}, 838 (1960).
\bibitem{nak} See, {\it e.g.}, N. Nakanishi, {\it Graph Theory and Feynman Integrals} (Gordon and Breach, New York, 1971).
\bibitem{bc} E. A. Bender and E. R. Canfield, J. Comb. Theory A {\bf 24}, 296 (1978).
\bibitem{corn96}   J. M. Cornwall, Phys. Rev. D {\bf 54}, 1814 (1996).
\bibitem{corn77} J. M. Cornwall, in {\it Deeper Pathways in High-Energy Physics}, edited by B. Kursonoglu, A. Perlmutter, and L. F. Scott (Plenum, New York, 1977), p. 683.
\bibitem{cornpa} J. M. Cornwall and J. Papavassiliou, Phys. Rev. D {\bf 40}, 3474 (1989).
\bibitem{lav} M. J. Lavelle, Phys. Rev. D {\bf 44}, R26 (1991).
\end{thebibliography}
\end{document}